# Plasmonic polaron in self-intercalated 1$T$-TiS$_2$

Byoung Ki Choi[1,2†*], Woojin Choi[3†], Zhiyu Tao[4,5†], Ji-Eun Lee[1,6,7], Sae Hee Ryu[1,8], Seungrok Mun[3], Hyobeom Lee[3], Kyoungree Park[3], Seha Lee[3], Hayoon Im[9], Yong Zhong[1,10,11], Hyejin Ryu[6], Min Jae Kim[9], Sue Hyeon Hwang[9], Xuetao Zhu[4,5], Jiandong Guo[4,5], Jong Mok Ok[9], Jaekwang Lee[9], Haeyong Kang[9], Sungkyun Park[9], Jonathan D. Denlinger[1], Heung-Sik Kim[3,12], Aaron Bostwick[1], Zhi-Xun Shen[10,11], Choongyu Hwang[9*], Sung-Kwan Mo[1*], and Jinwoong Hwang[3*]

[1]Advanced Light Source, Lawrence Berkeley National Laboratory, Berkeley, CA, USA.

[2]Technological Convergence Center, Korea Institute of Science and Technology, Seoul, Korea.

[3]Department of Physics and Institute of Quantum Convergence Technology, Kangwon National University, Chuncheon, Korea.

[4]Beijing National Laboratory for Condensed Matter Physics and Institute of Physics, Chinese Academy of Sciences, Beijing, China.

[5]School of Physical Sciences, University of Chinese Academy of Sciences, Beijing, China.

[6]Center for Semiconductor Technology, Korea Institute of Science and Technology, Seoul, Korea.

[7]Max Planck POSTECH Center for Complex Phase Materials, Pohang University of Science and Technology, Pohang 37673, Korea.

[8]Research Center for Beamline, Korea Basic Science Institute (KBSI), Daejeon, Korea.

[9]Department of Physics, Pusan National University, Busan, Korea.

[10]Geballe Laboratory for Advanced Materials, Department of Physics and Applied Physics, Stanford University, Stanford, CA, USA.

[11]Stanford Institute for Materials and Energy Sciences, SLAC National Accelerator Laboratory, Menlo Park, CA, USA.

[12]Department of Energy Engineering, Korea Institute of Energy Technology, Naju-si, Korea.

† equal contribution

* Corresponding authors: bkchoi@kist.re.kr, SKMo@lbl.gov, ckhwang@pusan.ac.kr, jwhwang@kangwon.ac.kr

KEYWORDS: 1$T$-TiS$_2$, Electron-plasmon coupling, Plasmonic polaron, ARPES, EELS, DFT

**Abstract**

Electron-boson coupling is central to a comprehensive understanding of the diverse physical phenomena emerging from many-body interactions. Yet less attention has been paid to how plasmons, collective bosonic modes of electron density oscillation, interact with conduction electrons and how external parameters can tune this interaction. Here, we present a clear display of composite quasiparticles stemming from electron-plasmon coupling, known as the plasmonic polaron, in self-intercalated 1$T$-TiS$_2$, by using angle-resolved photoemission spectroscopy (ARPES), high-resolution electron energy loss spectroscopy (HR-EELS) and first-principles calculations. The single particle spectral function exhibits a distinctive plasmon-loss satellite with the same characteristic energy scale determined by HR-EELS measurements. The bosonic energy scale of plasmonic polaron is tunable by controlling charge carrier density and temperature, distinguishing itself from conventional polarons arising from electron-phonon interactions. Furthermore, we find that the dielectric screening strongly affects the formation of the plasmonic polaron states. Our findings provide direct spectroscopic evidence of plasmonic polarons and establish self-intercalated layered materials as a promising platform for studying, controlling, and harnessing plasmonic interactions in quantum materials.

The interaction of electrons with collective bosonic modes in solids is of practical and fundamental interest in condensed matter physics. This interaction not only affects the transport properties of actual devices but also induces novel quantum phenomena. A prominent example is the formation of polarons, composite quasiparticles of electrons dressed in phonon clouds due to the strong electron-phonon interaction[1]. Polarons lead to significantly enhanced quasiparticle mass that modifies charge carrier transport properties, and play a critical role in superconducting and many-body phenomena in high-$T_c$ cuprates[2,3], manganites[4], two-dimensional (2D) electron gases in heterostructures[5], transition metal dichalcogenides[6,7], and superconducting monolayer FeSe[8]. Other notable examples include electron-magnon interactions in ferromagnetic metals[9] and electron interactions with antiferromagnetic spin fluctuations in iron-based superconductors[10].

Less studied electron-boson coupling in solids is the interaction between electrons and plasmons, quasiparticles of collective charge density oscillations. The electron-plasmon coupling can produce composite quasiparticles known as plasmarons or plasmonic polarons, characterized by distinctive plasmon-loss satellites in the energy spectrum[1,11], analogous to polarons from electron-phonon interaction[12,13]. Plasmonic polarons exhibit unique properties distinct from conventional phonon-induced polarons, which may offer advantages in specific applications. Notably, plasmonic polarons generally have higher characteristic energies and greater tunability than phononic polarons[11]. Since plasmons are excited with a characteristic frequency ($\omega_{\text{Plasmon}}$) related to the carrier density ($n$), plasmon energy can be effectively modulated by altering the charge carrier density[13,14] and temperature[15-17]. Furthermore, electron-plasmon coupling can significantly impact the low-energy properties of materials, such as enhancing quasiparticle mass and limiting charge carrier mobilities[2,18].

Despite the important role of electron-plasmon coupling in determining the electronic properties of solids, the plasmonic polaron has been scarcely investigated in real materials. The scarcity is primarily due to the relatively weak electron-plasmon coupling strength[18] and the challenge of achieving sufficiently high doping levels to drive a material system into a well-defined plasmonic regime while maintaining stoichiometry. To date, only a limited number of studies have reported this phenomenon, primarily focusing on heavily electron-doped oxide semiconductors and 2D materials all in their thin film forms[12,13,19], often involving external treatments such as post-annealing, surface irradiation[5,12], or extrinsic surface carrier doping[14,19]. Consequently, open questions still remain about whether there is a way to achieve an optimal charge carrier density to realize well-defined plasmonic polaron signatures in bulk materials without external treatments and how these signatures could be distinguished from those of other bosonic excitation modes.

In this paper, we report a direct observation of plasmonic polarons formed by electron-plasmon coupling in self-intercalated $1T$-TiS$_2$, using angle-resolved photoemission spectroscopy (ARPES) and high-resolution electron energy loss spectroscopy (HR-EELS) measurements. Our experimental results reveal that bulk $1T$-TiS$_2$ is a highly electron-doped semiconductor with a band gap of ~ 0.5 eV due to self-intercalated Ti atoms, which serve as a charge reservoir between $1T$-TiS$_2$ layers. The most pronounced feature in an ARPES band structure is a satellite band at ~ 0.2 eV below the conduction band minimum, typically observed in a polaronic system due to "shake-off" excitations in the photoemission process[20]. HR-EELS clearly distinguishes between phonon and plasmon excitation energies and uncovers that the bulk plasmon excitation corresponds to the polaronic energy obtained in ARPES. First-principle many-body calculations, including the calculated plasmon self-energy, successfully reproduce the satellite band, confirming the formation of plasmonic polarons in self-intercalated $1T$-TiS$_2$. Furthermore, we find that the energy of the satellites in self-intercalated 1$T$-TiS$_2$ is tunable by varying the charge carrier density, in agreement with the expectation within the plasmonic polaron picture, and that the modification of the band structure and dielectric environment by temperature strongly affects the formation of the plasmonic polaron. Our findings demonstrate that plasmonic polarons can be formed intrinsically in bulk materials via interstitial adatoms, providing a versatile material platform enabling a systematic study of the plasmonic polaron state.

## Results and Discussion

**Atomic and electronic structures of 1$T$-TiS$_2$ and Ti self-intercalated 1$T$-TiS$_2$**

Figure 1**a** illustrates the crystal structure of the layered octahedral 1$T$-TiS$_2$, which crystallizes in a CdI$_2$-type layered structure (*P-3m1*) with a hexagonal Ti sublayer sandwiched between two S sublayers. The three sublayers constitute a single layer of 1$T$-TiS$_2$, weakly bonded to adjacent layers by van der Waals forces. It is generally known that foreign atoms can easily enter in the van der Waals gaps in transition metal dichalcogenides (TMDCs)[21,22]. In the case of 1$T$-TiS$_2$, the self-intercalation of Ti atoms has been directly observed, and theoretical analysis suggests that excess Ti is energetically more favorable than S vacancies[23]. The extra atoms are randomly distributed between 1$T$-TiS$_2$ layers[23] and our Raman spectrum further supports the presence of intercalated Ti atoms, indicated by a broad shoulder peak (Sh) on the high-energy side of the A$_{1g}$ peak (Figure **S**1)[22,24]. Energy-dispersive X-ray spectroscopy (EDX) analysis (Figure **S**2) confirms the non-stoichiometric composition of TiS$_2$, which would give a chemical formula Ti$_{1.089}$S$_2$, assuming Ti intercalation. To account for the Ti intercalation in our calculation, we have adopted Hubbard $U$ correction to Ti 3$d$ orbitals and added excess electrons in 1$T$-TiS$_2$. Figure 1**b** presents a density functional theory (DFT) calculation for self-intercalated 1$T$-TiS$_2$, which accurately reproduces the ARPES data from the as-cleaved self-intercalated sample shown in Figure 1**c**. Using orbital-projected calculations via Wannier functions, we find that the valence band maximum (VBM) at the Γ point is primarily composed of S 3*p* orbitals, while the conduction band minimum (CBM) at the M point is dominated by Ti 3*d* orbitals (Figure S3).

While the overall ARPES band features are well matched with the DFT+$U$ calculations (green dashed lines in Figure 1c), the calculated band structures do not replicate the additional satellite feature observed ~0.24 eV below the Fermi energy ($E_F$) at the M point. This satellite band displays not only the well-defined "peak-dip-hump" structure but also a broad tail concentrated at higher binding energies at the M point (Figure 1d). The satellite feature exhibits momentum dependence, as evidenced by ARPES constant-energy maps (CEMs) (Figure S4). This satellite spectral feature resembles well-known polaronic states reported in previous ARPES studies[12,13,19,25,26]. The entire conduction band is scattered by a particular bosonic mode to higher binding energies, while the original band shape is preserved due to shake-off excitations of the bosonic mode during the photoemission process[20] (Figure 4b). Therefore, new spectral weight in the ARPES intensity map (Figure 1d) emerges due to these excitations below the quasiparticle band, separated by bosonic-excitation energy ($\hbar\Omega$). These characteristic features (the QP pocket at M and its satellite) were consistently observed across multiple cleaved samples and at different locations on each sample. While the self-intercalated Ti atoms are microscopically distributed at random (Ref. 23), our ARPES measurements—which average over a macroscopic beam spot (~50-100 μm)—reveal a highly uniform and reproducible electronic structure. This indicates that the self-intercalation is macroscopically homogeneous.

**Plasmonic polaron in the self-intercalated 1*T*-TiS$_2$**

In order to identify the bosonic mode coupled with electron, we have conducted low-energy excitation measurements using HR-EELS. Figure 2**a** shows a boson momentum ($q$)-resolved HR-EELS spectrum taken along the Γ−M direction at 35 K. The HR-EELS measurements reveal two characteristic collective excitations in self-intercalated 1$T$-TiS$_2$. A low-energy excitation mode within below ~50 meV originates from the lattice vibrations (phonon). It is well matched with the energy scale of the calculated phonon modes for self-intercalated 1$T$-TiS$_2$ (Figure **S5**). Another excited bosonic mode appears in the HR-EELS spectrum at ~200 meV. This high-energy mode is distinct from the phonon dispersion, with peak intensity strong only near the Γ and rapidly diminishing as the momentum increases (Figure **S6**). This characteristic feature is well described by the Landau damping in plasmon dispersion, where the plasmon decays into single particle electron-hole pair excitations[27,28]. In addition, the non-zero plasmon loss energy at $q = 0.0$ Å$^{-1}$ indicates that the experimentally obtained bosonic mode at ~200 meV originates from the bulk plasmon. Other possible collective modes, such as exciton or impurity states, can be reasonably ruled out due to the transient nature of exciton and momentum dependence of the observed satellites[29,30]. Given the similar energy scale of the ARPES satellite band at the M-point (Figure 1**d**), we conclude that the bosonic excitation coupled with electrons in self-intercalated 1$T$-TiS$_2$ is indeed the bulk plasmon rather than a phonon.

One of the key distinctions between conventional polarons, originating from the electron-phonon coupling, and plasmonic polarons is the charge carrier density dependence of their characteristic excitation energy. While the characteristic energy of phonon modes (e.g. Fröhlich polaron) remains independent of charge carrier density[11,12,20], the characteristic energy of the plasmonic polarons can be tuned by the carrier concentration[11,12,14]. To delineate the nature of the bosonic mode by tuning the charge carrier density further, we deposited Rb atoms on the surface of the as-cleaved self-intercalated 1$T$-TiS$_2$ at 20 K[31]. The overall spectral shift to higher binding energy reflects an increased Fermi energy due to electron doping (Figures. **2b-e** and Figure **S7,8**). More importantly, the characteristic energy ($\hbar\Omega$) determined by separation between the quasiparticle (QP) and satellite peaks at the M point increases with rising charge carrier density, from $197 \pm 0.8$ meV ($n = 1.37 \times 10^{21}$ cm$^{-3}$) to $214 \pm 1.2$ meV ($n = 4.43 \times 10^{21}$ cm$^{-3}$), as shown in Figure **2f**. The increased satellite energy supports the formation of plasmonic polarons and highlights the possibility of manipulating these composite particles by tuning the charge carrier density.

The formation of the plasmonic polaron is further supported by the density functional perturbation theory (DFPT). Figure **2g** shows the calculated spectral intensity, modified by the calculated plasmon self-energy using the cumulant expansion method (Figure **S9**). Our DFPT calculation accurately reproduces the spectral weight of the plasmon satellite state observed in the ARPES results (Figure **2h**), clearly demonstrating the formation of plasmonic polaron in self-intercalated 1$T$-TiS$_2$. We note that the spectral intensity in Figure **2h** only shows the first-order satellite feature, with no higher-order satellites observed, unlike other polaronic features reported in the literature[8,12-14,26]. Based on the energy distribution curve (EDC) profile of ARPES data taken at M point (Figure **1f**), the first-order satellite has ~8 times less intensity compared to the quasiparticle peak. Assuming that the second-order satellite weight is reduced by a similar factor, its intensity would be at the background level[12,13,19] (Figure S10).

**Temperature-dependent plasmonic dynamics in self-intercalated 1$T$-TiS$_2$**

Further insights into plasmonic polaron states can be gained by examining the temperature dependence of ARPES satellite features, as plasmon energy has a subtle yet clear temperature dependence[15-17]. Until now, the temperature-dependence study in plasmonic polaron systems has been considered challenging, as these states are typically induced by light irradiation, dosing of the adatoms on the surface, or ferromagnetic transition[12,13,19,26], which are not sustainable at high temperatures. In contrast, the self-intercalated 1$T$-TiS$_2$, with its robust interstitial Ti atoms, is well-suited for investigating temperature-dependent plasmonic states.

Figure 3**a** presents the temperature-dependent HR-EELS spectra for self-intercalated 1$T$-TiS$_2$ taken at $q = 0.0$ Å$^{-1}$. The obtained HR-EELS spectra exhibit three features with increasing temperature: (i) a gradual shift of the bulk plasmon resonance peak towards lower loss energy, (ii) a reduction in the plasmon peak intensity, and (iii) an increase in the full width at half maximum (FWHM). These behaviors correspond well to plasmon damping due to increased electron-electron and electron-phonon scatterings at elevated temperatures[16,27,32]. The plasmon damping effects are also captured in the plasmonic polaronic state in ARPES results. Figure 3**b** shows the temperature-dependent ARPES spectra taken at the M point, where the plasmonic polaron state emerges (Figure 1**d**). With increasing temperature, EDCs clearly show a reduction in spectral weight of QP peaks along with suppression of the plasmon satellite peak (Figure 3**b** and **S**11). To characterize the temperature-dependent plasmon damping, we plot the FWHMs of both HR-EELS and ARPES spectra in Figure 3**c**. The FWHMs of the plasmon state in both spectra increase dramatically (~50 meV), while QP FWHM increases more moderately (~12 meV). As temperature rises, plasmonic polarons become rapidly unstable compared to the QP peak due to plasmon damping from enhanced lattice vibrations and thermal excitations, leading to the dissolution of polaronic states at high temperatures[16,27,32]. As a result, the features

associated with plasmonic polarons weaken, and FHWM increases with reduced QP intensity due to increased inelastic scattering, consistent with HR-EELS observations. We further notice that both QP and satellites peaks in ARPES spectra shift to higher binding energies with increasing temperature, but the satellite peak moves less, reducing the separation between the QP and satellite peaks (Figure 3**b** (inset) and **S**12).

In Figure 3**d**, the bosonic energies extracted from ARPES (red points) and HR-EELS (green points) both decrease with increasing temperatures. To investigate the temperature dependence of electron-plasmon coupling, we extracted the charge carrier density ($n$) and the effective mass ($m^*$) of the electron pocket at the M point from ARPES data. Our analysis reveals a gradual increase in $n$ and a decrease in $m^*$ with rising temperature (Figure **S**13). Considering the relation for bulk plasmon energy, $\hbar\Omega \propto \sqrt{n/\epsilon m^*}$, the temperature dependent behavior of $\sqrt{n/m^*}$ contrasts sharply with the observed reduction in plasmon energy at higher temperatures (Figures 3**d** and 4**a**). This discrepancy suggests that the dielectric constant ($\epsilon$) must increase with temperature to reconcile the observed trends in $n$ and $m^*$, indicating that $\epsilon$ plays a critical role in modulating plasmon energy as a function of temperature (Figure 4**a**). To extract $1/\epsilon$, we have used the plasmon energy relation of $\hbar\Omega \propto \sqrt{n/\epsilon m^*}$, where the plasmon energy ($\hbar\Omega$) and $\sqrt{n/m^*}$ were extracted from ARPES results (Figs. 3**d**, **S**12 and **S**14). As $\epsilon$ can be influenced by changes in $n$ and crystal structure[33,34], the observed increase in $n$ partially contributes to the enhancement of $\epsilon$ at elevated temperatures. The minor discrepancies between plasmon energies measured by ARPES and HR-EELS (Fig. 3**d**) can be attributed to the different mechanisms inherent to these techniques[32,35-37].

Additionally, temperature-dependent in-plane tensile stress of 1$T$-TiS$_2$ further contributes to the overall temperature dependence of plasmon energies. Temperature-dependent ARPES data show an unexpected band shift at the Γ point, where the S-induced hole band shifts to higher binding energy with increasing temperature. This band shift is well reproduced by DFT calculations under an in-plane tensile stress of ~1.3%, consistent with Raman spectroscopy measurements[22] (Figures **S**15 and **S**16). These findings suggest that observed tensile stress is an additional factor influencing the temperature evolution of plasmonic energy. Therefore, a comprehensive understanding of plasmonic polarons requires the combined use of ARPES, HR-EELS, and theoretical calculations.

It is fascinating to notice that similar satellite features in ARPES spectra have been discussed in connection with the enhancement of superconductivity, e.g., at the interface of single-layer FeSe and $SrTiO_3$[8,38] and Cr-intercalated $ZrTe_2$[39]. While BCS theory may be extended to a general bosonic field including plasmon[1], it is not yet clear whether plasmonic polaron may play a similar role as the phononic polarons in creating and enhancing superconductivity. It would be interesting to see whether it is possible to find similar satellite features through the intercalation of charge-donating atoms in the van der Waals gap of superconducting TMDCs such as $NbSe_2$[40,41], and how the superconductivity is affected when plasmonic polarons are formed in these systems.

In summary, our combined investigation with ARPES, HR-EELS, and DFT on the exotic bosonic coupling in self-intercalated $1T$-$TiS_2$ unambiguously confirms the presence of plasmonic polarons. This result establishes self-intercalated material as a promising platform to utilize the plasmonic polaron with a highly tunable electronic structure that can be modified by carrier density and temperature. Our study demonstrates that TMDCs are ideal materials for the creation, study, and control of plasmonic polarons in bulk, owing to their propensity to incorporate additional transition metal atoms into the van der Waals gap between layers[42-44]. Moreover, its layered structure and intrinsic plasmonic nature make self-intercalated 1$T$-TiS$_2$ advantageous for constructing van der Waals heterostructures, which is promising in exploiting plasmonic physics, such as realizing high-temperature superconductors through hybrid phonon-plasmon coupling and opening new opportunities for plasmonic applications.

**Methods**

**Sample and ARPES measurement**

The self-intercalated 1$T$-TiS$_2$ single crystals were purchased from HQ Graphene. The EDX measurements were performed at the Korea Institute of Science and Technology (KIST) to determine the ratio of atomic weights. ARPES data were taken at Beamlines 10.0.1 and 4.0.3, Advanced Light Source, Lawrence Berkeley National Laboratory, using Scienta R4000 and R8000 analyzers, respectively. The base pressure was better than $4 \times 10^{-11}$ Torr. The photon energy was set at 52 eV for $p$-polarizations with energy and angular resolution of 18–25 meV and 0.1°, respectively. To achieve high-quality ARPES data of self-intercalated TiS$_2$, the samples were cleaved at low temperatures (~20 K). Rubidium (Rb) deposition was carried out by *in-situ* evaporation of Rb on the sample surfaces using a commercial SAES getter source mounted in the analysis chamber.

**HR-EELS measurements**

The 1$T$-TiS$_2$ single crystals were also examined with an HR-EELS system equipped with an Ibach-type ELS5000 monochromator from LK technologies and an Scienta R4000 analyzer[45]

HR-EELS measurements were performed at temperatures ranging from 35 K to 291 K with a primary electron beam energy of 110 eV.

**Density Functional Theory Calculations**

All DFT and DFPT calculations for 1$T$-TiS$_2$ (space group $P\bar{3}m1$) were performed using the Quantum ESPRESSO[46] package. We adopted the Hubbard corrections (DFT+$U$) for hopping mechanism of intercalation with an effective onsite Coulomb parameter $U$ = 3.25 eV for Ti (3$d$). The generalized gradient approximation (GGA) of Perdew, Burke and Ernzerhof for solids (PBEsol) exchange-correlation functionals and norm-conserving pseudopotentials were used. We employed a plane wave kinetic energy cutoff of 50 Ry (680 eV), kinetic energy cutoff for charge density and potential for norm-conserving pseudopotential of 500 Ry, and 12 × 12 × 12 Monkhorst-Pack mesh to sample the Brillouin zone. Estimated self-consistent field accuracy is equal to 1 × 10$^{-12}$ Ry. The phonon dispersions were obtained with a 4 × 4 × 4 supercell. To illustrate the real space orbital and orbital projected bands, we used Wannier function disentanglement with Wannier90[47]. We used cumulant expansion function in EPW code[48] to obtain spectral functions with electron-phonon and electron-plasmon self-energy. The retarded electron self-energy is driven by Raleigh-Schrödinger perturbation theory:

$$\int \frac{dq}{\Omega_{BZ}} \sum_{m\nu} |g_{mn}^{\nu}(k,q)|^2 \left[ \frac{n(\omega_{q\nu}) + f(\epsilon_{mk+q})}{\epsilon_{nk} - \epsilon_{mk+q} - \omega_{q\nu} - i\eta} + \frac{n(\omega_{q\nu}) + 1 - f(\epsilon_{mk+q})}{\epsilon_{nk} - \epsilon_{mk+q} + \omega_{q\nu} - i\eta} \right].$$

Here, $\epsilon_{mk+q}$ is electron energy, $n(\omega_{q\nu})$ is the Bose occupation factors, $f(\epsilon_{mk+q})$ is Fermi-Dirac occupation factors, $\eta$ is infinitesimal, $\omega_{q\nu}$ is plasmon/phonon frequency. The $g_{mn}^{\nu}(k,q)$ represents matrix elements between the initial state $\psi_{nk}$ and the final state $\psi_{mk+q}$ for electron-phonon and electron-plasmon coupling[49] given by

$$g_{mn}^{\nu}(k, q) = \langle\psi_{mk+q}|\partial_{q\nu}V|\psi_{nk}\rangle, \text{ and } g_{mn}^{\nu}(k, q) = \left[\frac{\varepsilon_0 \Omega}{e^2 \hbar}\frac{\partial\epsilon(q,\omega)}{\partial\omega}\right]_{\omega_q}^{-1/2} \frac{1}{|q|}\langle\psi_{mk+q}|e^{iq\cdot r}|\psi_{nk}\rangle$$

, respectively, with $\partial_{q\nu}V$ the derivative of the self-consistent potential associated with a phonon of wavevector q, ν branch index, Ω the volume of one unit cell.

**Acknowledgements**

This research used resources of the Advanced Light Source, which is a DOE Office of Science User Facility under contract no. DE-AC02-05CH11231. BKC and JEL were supported in part by an ALS Collaborative Postdoctoral Fellowship. The work at the SIMES/Stanford is supported by the U.S. Department of Energy, Office of Basic Energy Sciences, Division of Materials Sciences and Engineering, under Contract No. DE-AC02-76SF00515. J.H. acknowledges financial support from the National Research Foundation of Korea (NRF) grant funded by the Korean government (MSIT) (RS-2023-00280346), the GRDC (Global Research Development Center) Cooperative Hub Program through the NRF funded by the Ministry of Science and ICT (MIST) (RS-2023-00258359), and Semiconductor R&D Support Project through the Gangwon Technopark (GWTP) funded by Gangwon Province (No. GWTP 2023-027). C.H. acknowledges financial support from NRF grant funded by the Ministry of Science and ICT (RS-2025-25453111, RS-2023-00221154, RS-2022-NR068223), the KBSI (NFEC) grant funded by the Ministry of Education (RS-2024-0043534, RS-2021-NF000587), and PNU-RENovation(2023-2024). H. R. and BKC acknowledges financial support from the KIST Institutional Program (2E33581, 2V10511). H. R. acknowledges financial support from the NRF (2020R1A5A1016518) and NST (GTL24041-000) grants by the MSIT.

**Author contributions**

B.K.C., J.W.H., C.H. and S.-K.M. initiated and conceived the research. J.W.H., B.K.C. and W.J.C. performed the ARPES measurements with the help from S.H.R., J.-E.L., H.B.L., K.R.P., S.H.L., J.D.D, H.J.R and Y.Z. and under the supervision of S.-K.M., A.B., Z.-X.S. and C.H. J.W.H. and B.K.C. analyzed the ARPES data. S.R.M., H.-S.K., J.L., and B.K.C. performed DFT calculations and theoretical analyses. Z.T. performed the HR-EELS measurements under the supervision of X.Z. and J.G., and S.H.H., M.J.K., J.M.O., S.P., H.K., and C.H. analyzed the HR-EELS data. B.K.C., W.J.C., J.W.H., C.H., Y.Z. and S.-K.M. wrote the manuscript with the help from all authors. All authors contributed to the scientific discussion.

**Competing interests**

The authors declare no competing interests.

**Data availability**

The data that support the plots within this paper and other findings of this study are available from Jinwoong Hwang (jwhwang@kangwon.ac.kr), who will respond to requests for data upon reasonable request.

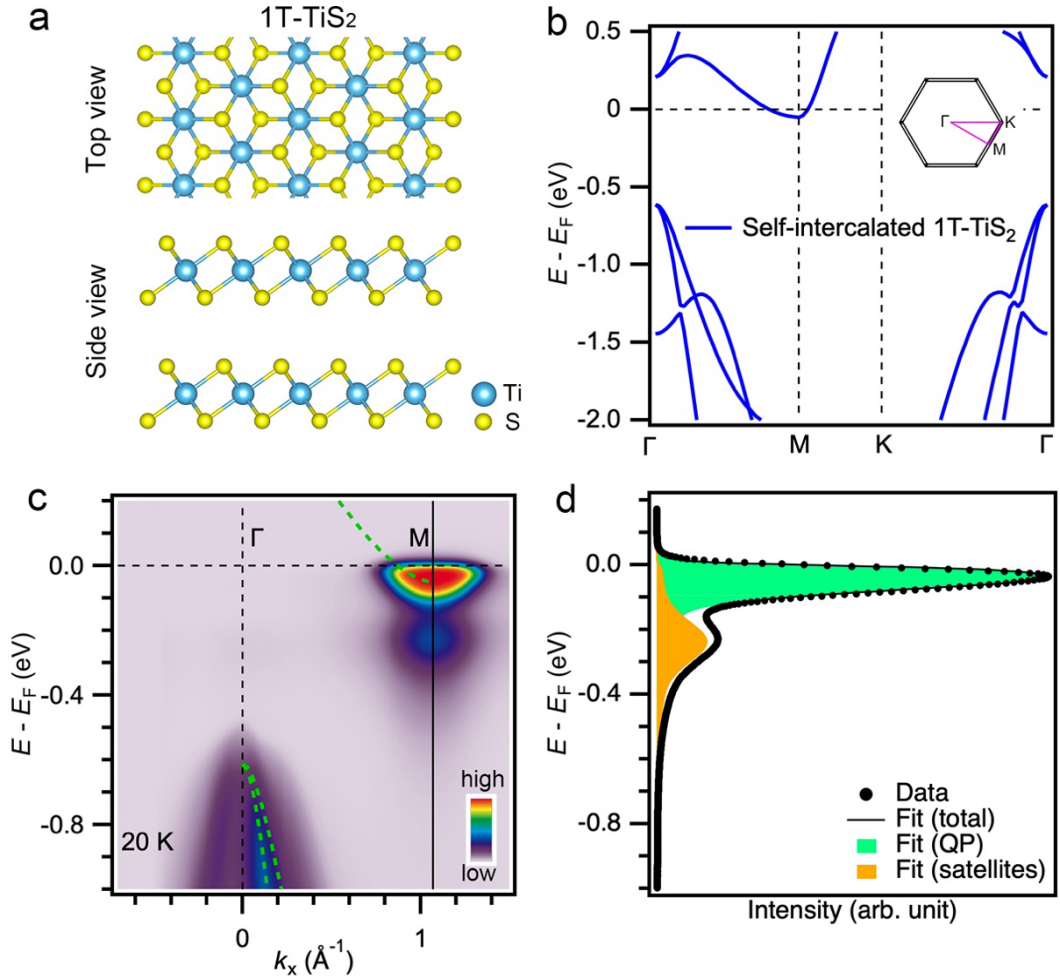

**Figure 1 Atomic and electronic structures of 1$T$-TiS$_2$ and Ti self-intercalated 1$T$-TiS$_2$. a**, Top- and side-view schematics of TiS$_2$ with 1$T$ structure. The sky-blue and yellow balls represent the Ti and S atoms, respectively. **b**, DFT calculated electronic structures of self-intercalated TiS$_2$ with 1$T$ structure. Hubbard $U$ correction to Ti 3$d$ orbitals and added excess electrons are adopted in 1$T$-TiS$_2$ to mimic the electronic structure of the self-intercalated 1$T$-TiS$_2$. The amount of doped electron is one electron per 8 unit-cells. (Inset: The Brillouin zone of 1$T$-TiS$_2$ with high-symmetry points.) **c**, ARPES spectra of the as-cleaved self-intercalated TiS$_2$ (Ti$_{1.089}$S$_2$) along the Γ-M direction for Ti-intercalated TiS$_2$ with the $p$-polarization. Calculated bands are overlapped with green dashed line. **d**, Energy distribution curve (EDC) taken at $k_x$ = 1.07 Å$^{-1}$ (M point). Green and orange areas indicate quasiparticle peak and its satellite peak, respectively.

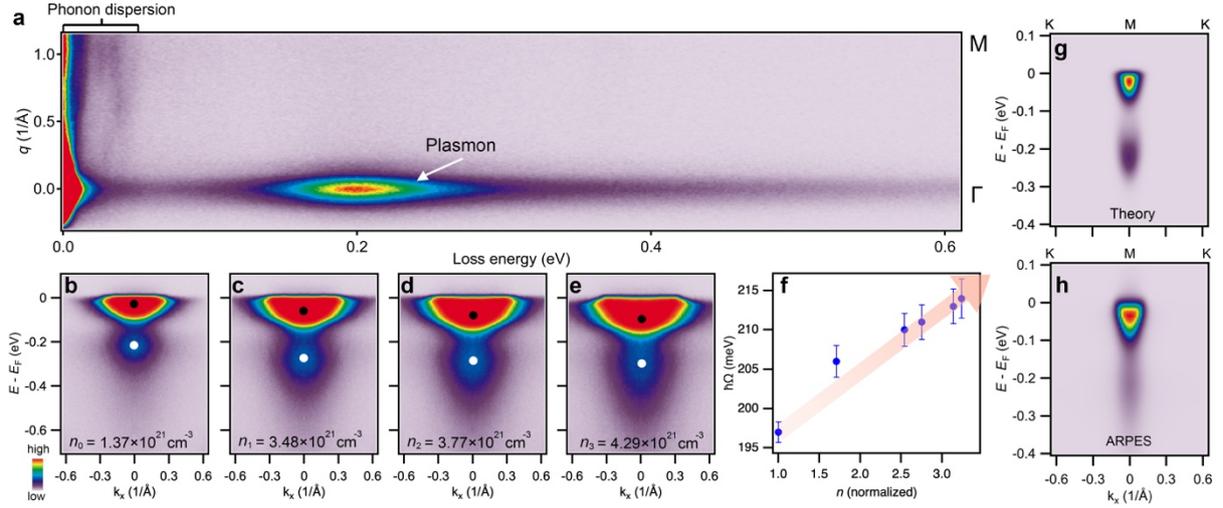

**Figure 2 Plasmonic polaron in the self-intercalated 1$T$-TiS$_2$**. **a**, Boson momentum-resolved HR-EELS of the self-intercalated 1$T$-TiS$_2$ at 35 K. $q$ indicates boson momentum transfer. **b-e**, ARPES intensity maps of self-intercalated 1$T$-TiS$_2$ sample after additional electron doping via in-situ Rb deposition. **f**, The evolution of bosonic energy extracted from ARPES data (**b-e**) as a function of $n$. **g-h**, DFPT calculated spectral function and ARPES spectra along the K−M−K direction for self-intercalated 1$T$-TiS$_2$. The calculated spectral function is post-processed with Gaussian broadening and Fermi-Dirac statistic (Figure **S**7).

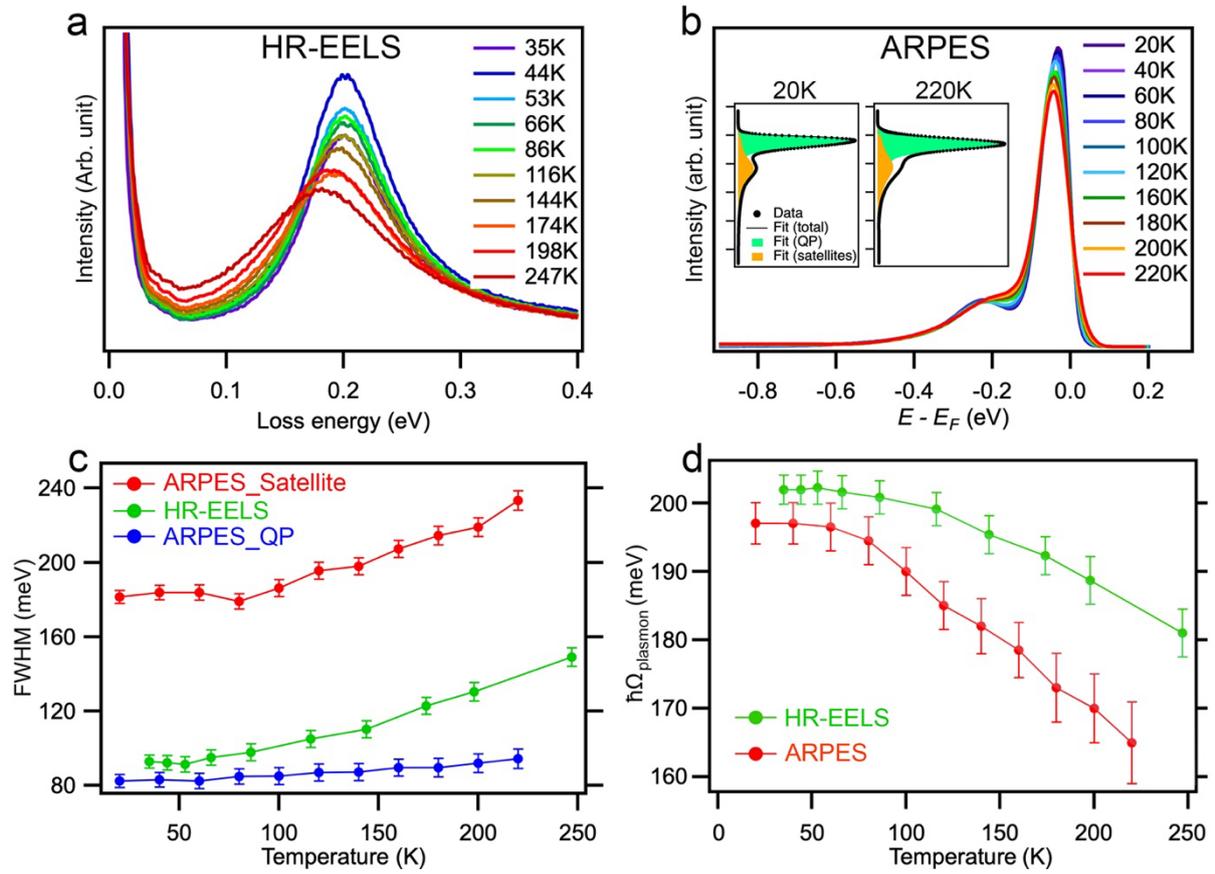

**Figure 3 Temperature-dependent plasmonic dynamics in self-intercalated 1*T*-TiS$_2$**. **a**, Temperature-dependent EDCs obtained from EELS at $q$ = 0.0 Å$^{-1}$. **b**, Temperature-dependent EDCs obtained from ARPES at the M point. **c**, Temperature-dependent FWHMs obtained from satellite and QP peaks in ARPES (red and blue points) and HR-EELS (green points), respectively. **d**, Evolution of bosonic energy obtained from ARPES (red points) and HR-EELS (green points) as a function of temperature.

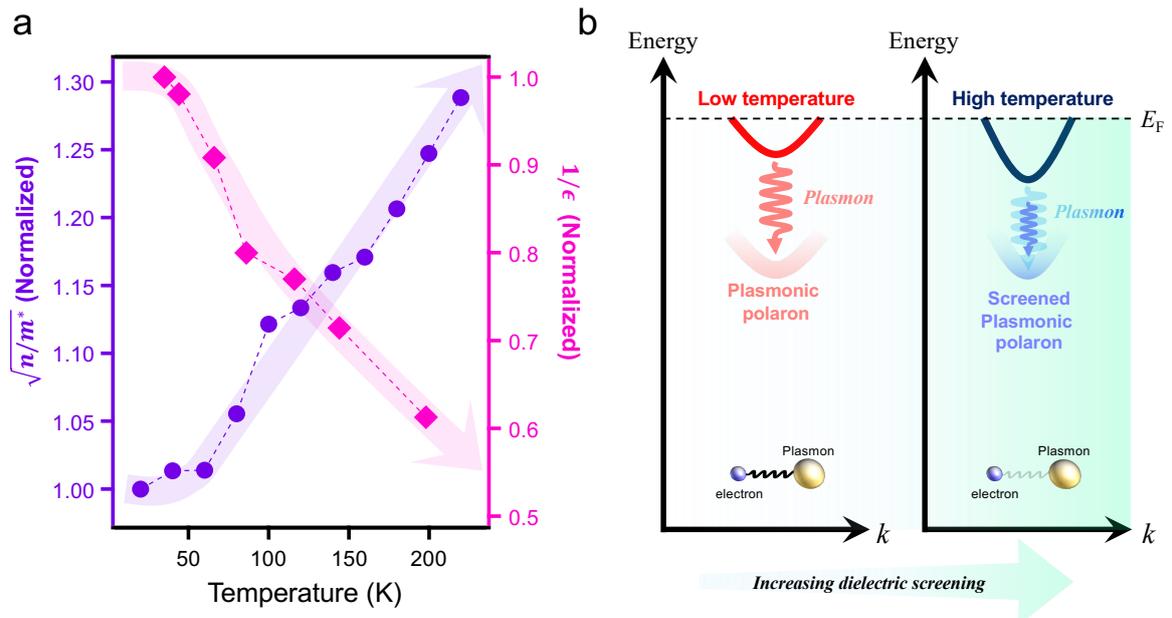

**Figure 4 Temperature-dependent many-body characteristic of self-intercalated 1$T$-TiS$_2$.**
**a**, Normalized values of $\sqrt{n/m^*}$ (purple circles) and $1/\epsilon$ (pink rhombus) as a function of temperature, respectively. The normalizations were based on the value of 20 K. **b**, Schematic illustrating the formation of the plasmonic polaron in self-intercalated 1$T$-TiS$_2$ at low and high temperatures. Light green background indicates the increase of the dielectric screening.